
\input harvmac.tex
\noblackbox
\let\includefigures=\iftrue
\newfam\black
\includefigures

\input epsf
\def\figin{\epsfcheck\figin}\def\figins{\epsfcheck\figins}
\def\epsfcheck{\ifx\epsfbox\UnDeFiNeD
\message{(NO epsf.tex, FIGURES WILL BE IGNORED)}
\gdef\figin##1{\vskip2in}\gdef\figins##1{\hskip.5in}
\else\message{(FIGURES WILL BE INCLUDED)}%
\gdef\figin##1{##1}\gdef\figins##1{##1}\fi}
\def\DefWarn#1{}
\def\figinsert{\goodbreak\midinsert}
\def\ifig#1#2#3{\DefWarn#1\xdef#1{fig.~\the\figno}
\writedef{#1\leftbracket fig.\noexpand~\the\figno}%
\figinsert\figin{\centerline{#3}}\medskip
\centerline{\vbox{\baselineskip12pt
\advance\hsize by -1truein\noindent
\footnotefont{\bf Fig.~\the\figno:} #2}}
\bigskip\endinsert\global\advance\figno by1}
\else
\def\ifig#1#2#3{\xdef#1{fig.~\the\figno}
\writedef{#1\leftbracket fig.\noexpand~\the\figno}%
\centerline{\vbox{\baselineskip12pt
\footnotefont{\bf Fig.~\the\figno:} #2}}
\global\advance\figno by1}
\fi

\input xy
\xyoption{all}
\font\cmss=cmss10 \font\cmsss=cmss10 at 7pt
\def\inbar{\,\vrule height1.5ex width.4pt depth0pt}
\def\IC{{\relax\hbox{$\inbar\kern-.3em{\rm C}$}}}
\def\IP{{\relax{\rm I\kern-.18em P}}}
\def\IF{{\relax{\rm I\kern-.18em F}}}
\def\IZ{\relax\ifmmode\mathchoice
{\hbox{\cmss Z\kern-.4em Z}}{\hbox{\cmss Z\kern-.4em Z}}
{\lower.9pt\hbox{\cmsss Z\kern-.4em Z}}
{\lower1.2pt\hbox{\cmsss Z\kern-.4em Z}}\else{\cmss Z\kern-.4em
Z}\fi}
\def\td{{\hbox{Td}}}
\def\co{{\cal O}}
\def\cl{{\cal L}}
\def\cp{{\cal P}}
\def\cm{{\cal M}}
\def\cn{{\cal N}}
\def\cs{{\cal S}}

\def\ch{{\hbox{ch}}}
\def\wv{{\widetilde V}}
\def\wb{{\widetilde B}}
\def\wz{{\widetilde Z}}

\def\wf{{\widetilde F}}
\def\wal{{\widetilde \alpha}}
\def\wsig{{\widetilde \sigma}}

\def\wb{{\widetilde B}}

\def\ra{{\rightarrow}}
\def\ip{{\IP}}
\def\vp{{V^\prime}}
\def\vpp{{V^{\prime\prime}}}

\def\CYt{Calabi--Yau threefold}
\def\CYf{Calabi--Yau fourfold}
\def\ie{{\it i.e.,\/}}
\Title{\vbox{\baselineskip12pt\hbox{hep-th/0009228}
\hbox{IASSNS-HEP-00/73} 
\hbox{UTTG-04/00}
\hbox{DUKE-CGTP-00/19}}}
{\vbox{
\centerline{Codimension-Three Bundle Singularities}
\centerline{in F-Theory}}}
{\vbox{\centerline {Philip Candelas$^1$, Duiliu-Emanuel Diaconescu$^2$, 
Bogdan Florea$^{1,3}$,}}}
{\vbox{\centerline{David R. Morrison$^{2,4}$, and Govindan Rajesh$^2$}}}
\bigskip
\bigskip
\medskip
\centerline{$^1$ {\it Mathematical Institute, University of Oxford,}}
\centerline{{\it 24-29 St. Giles', Oxford OX1 3LB, England}} 
\centerline{{$^2$ \it School of Natural Sciences, Institute for Advanced
Study,}} 
\centerline{{\it Einstein Drive, Princeton, NJ 08540, USA}} 
\centerline{$^3$ {\it Theory Group, Department of Physics, 
University of Texas at Austin,}} 
\centerline{{\it Austin, TX, 78712, USA}}
\centerline{{$^4$ \it Center for Geometry and Theoretical Physics,
Duke University,}}
\centerline{{\it Durham, NC 27708, USA}} 
\bigskip
\bigskip
\bigskip
\noindent
We study new nonperturbative phenomena in $N=1$ heterotic string vacua
corresponding to pointlike bundle singularities in codimension three. 
These degenerations result in new four-dimensional infrared physics
characterized by light solitonic states whose origin is explained 
in the dual F-theory model. We also show that such phenomena appear
generically in $E_7 \to E_6$ Higgsing and describe in detail the 
corresponding bundle transition.

\Date{September 2000}
\lref\FMW{R. Friedman, J.W. Morgan, and E. Witten, ``Vector Bundles 
over Elliptic Fibrations,'' J. Algebraic Geom. {\bf 8} (1999) 279, 
alg-geom/9709029.}

\newsec{Introduction and Overview}

A remarkable achievement of string theory in recent years 
consists of 
understanding various nonperturbative effects associated to the 
breakdown of worldsheet conformal field theory. An example which has 
received 
much attention in the literature is the small instanton 
singularity 
in heterotic string theories 
\nref\hetI{E. Witten, ``Small Instantons in String Theory,''
Nucl. Phys. {\bf B480} (1996) 213, hep-th/9511030.}%
\hetI.
These singularities occur in the context of heterotic string 
theories 
compactified on a K3 surface and are associated to the simplest 
pointlike 
degenerations of the background gauge bundles. 
Such degenerations have been shown to result in nonperturbative 
effects in six dimensions
which can be understood  either in terms of D-brane physics
\nref\dg{M.R. Douglas and G. Moore, 
``D-branes, Quivers, and ALE Instantons,'' hep-th/9603167.}%
\nref\I{K. Intriligator, ``RG Fixed Points in Six Dimensions via 
Branes at Orbifold Singularities,'' Nucl. Phys. {\bf B496} (1997) 177,
hep-th/9702038.}%
\nref\BI{J. Blum and K. Intriligator, 
``Consistency Conditions for Branes at Orbifold Singularities,'' 
Nucl. Phys. {\bf B506} (1997) 223, hep-th/9705030.}%
\nref\BItwo{J. Blum and K. Intriligator, 
``New Phases of String Theory and 6d RG Fixed Points via Branes at 
Orbifold Singularities,'' Nucl. Phys. {\bf B506} (1997) 199, 
hep-th/9705044.}%
\refs{\hetI\dg\I\BI{--}\BItwo},
or more generally, from the point of view of F-theory 
\nref\mvtwo{D.R. Morrison and C. Vafa, 
``Compactifications of F-Theory on 
Calabi--Yau threefolds -- II,'' Nucl.Phys. {\bf B476} (1996)
437, hep-th/9603161.}%
\nref\ag{P. Aspinwall and M. Gross, ``The $SO(32)$ Heterotic 
String on a K3 Surface,'' Phys. Lett. {\bf B387} (1996) 735, 
hep-th/9605131.}%
\nref\A{P. Aspinwall, ``Point-like Instantons and the 
Spin(32)/Z2 Heterotic String,'' Nucl. Phys. {\bf B496} (1997) 149,
hep-th/9612108.}%
\nref\am{P. Aspinwall and D.R. Morrison, 
``Point-like Instantons on K3 Orbifolds,'' 
Nucl. Phys. {\bf B503} (1997) 533, hep-th/9705104.}%
\refs{\mvtwo\ag\A{--}\am}.
Various nonperturbative aspects of four dimensional heterotic 
strings 
have also been studied in detail 
\nref\M{P. Mayr, ``Mirror Symmetry, N=1 Superpotentials and 
Tensionless Strings on Calabi--Yau Four-Folds,'' Nucl. Phys. {\bf B494}
(1997) 489, hep-th/9610162.}%
\nref\low{A. Lukas, B.A. Ovrut, and D. Waldram, 
``Non-standard embedding and five-branes in heterotic M-Theory,''
Phys.Rev. {\bf D59} (1999) 106005,  hep-th/9808101.}%
\nref\dlow{R. Donagi, A. Lukas, B.A. Ovrut, and D. Waldram,
``Non-Perturbative Vacua and Particle Physics in M-Theory,'' 
JHEP {\bf 05} (1999) 018, hep-th/9811168.}%
\nref\toric{
G. Rajesh, ``Toric Geometry and 
F-theory/Heterotic Duality in Four Dimensions,'' 
JHEP {\bf 12}
(1998), hep-th/9811240.}%
\nref\mirror{ 
P. Berglund and P. Mayr, ``Heterotic String/F-theory Duality from
Mirror Symmetry,'' ATMP {\bf 2} (1998) 1307, hep-th/9811217.}%
\nref\dlowB{R. Donagi, A. Lukas, B.A. Ovrut, and D. Waldram,
``Holomorphic Vector Bundles and Non-Perturbative Vacua in M-Theory,'' 
JHEP {\bf 06} (1999) 034, hep-th/9901009.}%
\nref\dr{D.-E. Diaconescu and G. Rajesh, 
``Geometrical Aspects of Fivebranes in Heterotic/F-Theory 
and F-theory/Heterotic Duality in Four Dimensions,'' JHEP {\bf 06}
(1999), hep-th/9903104.}%
\nref\dow{R. Donagi, B.A. Ovrut, and D. Waldram,
``Moduli Spaces of Fivebranes on Elliptic Calabi--Yau Threefolds,'' 
JHEP {\bf 11} (1999) 030, hep-th/9904054.}%
\nref\bpp{B.A. Ovrut, T. Pantev, and J. Park, 
``Small Instanton Transitions in Heterotic M-Theory,''  
JHEP {\bf 05} (2000) 045, hep-th/0001133.}%
\nref\ggo{A. Grassi, Z. Guralnik, and B.A. Ovrut, 
``Five-Brane BPS States in Heterotic M-Theory,'' hep-th/0005121.}%
\refs{\M\low\dlow\toric\mirror\dlowB\dr\dow\bpp{--}\ggo}.
The common feature of all these effects is that they can be 
ultimately 
related to the six dimensional small instanton singularity by 
an adiabatic
argument. They have been accordingly interpreted in terms of 
heterotic 
fivebranes wrapping holomorphic curves in the Calabi--Yau 
threefold. 
From a mathematical point of view, such CFT singularities 
correspond 
to codimension-two bundle degenerations, precisely as in 
the six dimensional 
situation. 

In this paper we consider a new class of nonperturbative effects 
specific 
to $N=1$ heterotic compactifications on Calabi--Yau threefolds. 
The singularities treated in the 
present work are 
qualitatively new, being associated to 
{\it codimension-three bundle
degenerations}. This is a novel class of degenerations which have 
not been studied in physics so far and which are specific to four 
dimensional 
$N=1$ compactifications. As such, we expect qualitatively new 
infrared effects in four dimensional $N=1$ theories which will be 
discussed below. 

The main tool for analyzing these singularities is heterotic/F-theory 
duality which encodes the bundle data in the geometry of a 
singular 
Calabi--Yau fourfold. This gives a pure geometric 
interpretation of the 
perturbative heterotic spectrum and determines at the 
same time the 
nonperturbative massless spectrum associated to CFT singularities.
Generically, pointlike bundle singularities are expected to result 
in a certain type of spacetime defect where the nonperturbative 
degrees of freedom are localized. While this is also the case 
here, 
the nature of the resulting defect is very hard to understand. 
This is caused by our poor understanding of codimension-three 
degenerations of solutions to the Donaldson--Uhlenbeck--Yau 
equation. 
In particular, no explicit throat-like supergravity solution 
is known 
in this case.

In order to gain some insight into the nature of these singularities, 
it may be helpful to highlight the most important physical aspects
by comparison with the small instanton transition. 
The $E_8$ bundle acquires pointlike degenerations which can be 
regarded 
as three dimensional defects filling space-time. However, 
there are no such 
stable excitations in the bulk M-theory, therefore such a 
defect is 
effectively stuck to the nine dimensional wall. This fact 
makes its 
physical properties quite obscure since it is not clear 
how to identify 
the light states governing the dynamics.

The F-theory picture is however more explicit. 
As expected, the bundle singularities 
correspond 
to special points on the F-theory base where the elliptic 
fibration 
develops certain non-generic singularities. These are 
superficially 
similar to the singularities occurring in the F-theory 
presentation of $E_8$ small instantons. So one might think by
analogy that each such defect would correspond to a blowup 
of the 
three complex dimensional base. In fact, this is not the 
case since it will be shown in section two that the 
smooth fourfold obtained by blowing up the base is not 
Calabi--Yau. 
This is in good agreement with the absence of a 
`Coulomb branch' noted 
previously (since the size of the exceptional divisor 
would be related 
to a displacement of the defect in the M theory bulk, 
which is forbidden). 

Quite remarkably, it turns out that in the present case, there 
exist Calabi--Yau resolutions involving only fiber blowups.  
Recall that the resolution 
of the typical ADE singular fibers occurring in F-theory 
consists of a 
chain of two-spheres with specific intersection numbers in 
agreement with 
the corresponding Dynkin diagram. On top of each point in 
the base we have 
generically such a collection of spheres. 
A careful analysis reveals the fact that above the special 
singular points 
the resolved fiber contains an entire {\it complex\ 
surface,} \ie\ a manifold 
of dimension four rather than a collection of two-spheres. 
Even more surprising is the fact that the occurrence of this 
surface 
is basically automatic; no extra blowups are necessary and 
there are 
no extra generators of the K\"ahler cone. 

This has interesting consequences for physics, which are easier to 
understand by compactifying the four dimensional F-theory model 
down 
to three dimensions on a circle of radius $R$. 
According to standard duality, this is equivalent 
to M theory on the resolved fourfold, the size of the elliptic 
fiber
being proportional to  $1/R$. The presence of the surface in 
the fiber 
results in new 
light degrees of freedom in the low energy spectrum. We can 
have a 
string corresponding to the M fivebrane wrapped on the 
surface $S$ and  a tower of particle states arising 
by wrapping membranes on holomorphic curves in $S$. 
We regard the nonperturbative massless excitations as a sign 
of a singularity in the heterotic $(0,2)$ CFT. However, at the 
present stage it is very hard to get more insight into the low 
energy dynamics. 

After this outline of the physics, let us describe next the 
precise context in which such singularities may be encountered.
It is a common fact in string theory that singularities of 
various sorts are associated to phase transition between 
string vacua. 
As discussed in more detail later, it turns out that the 
pointlike singularities considered here appear generically 
in the context of the Higgs phenomenon in F-theory. 
More explicitly,  we consider a typical $E_6\rightarrow E_7$ 
transition corresponding to a family of singular 
Calabi--Yau fourfolds with generic fiber singularity $E_6$ which 
is enhanced 
to $E_7$ along a subspace of the moduli space. Technically, such 
an 
enhancement is realized by setting to zero certain parameters 
of the 
Weierstrass model. When this apparently simple transition is 
studied 
in detail one notices the presence of extra codimension-three 
singularities and the nonperturbative phenomena described above. 

We can get a new perspective on this transition by making use of
the spectral cover construction of Friedman, Morgan, and Witten \FMW.
At generic points in the moduli space we have a smooth holomorphic
bundle of rank three. At the transition point, this bundle degenerates 
in a 
controlled manner to a singular object which is technically a 
coherent 
sheaf. Coherent sheaves have made their appearance in a number 
of places
in physics. For example, in the linear sigma model approach 
to $(0,2)$ modes 
\nref\linsigm{E. Witten, ``Phases of $N=2$ Theories in Two
Dimensions,'' Nucl. Phys. {\bf B403} (1993) 159, hep-th/9301042.}%
\linsigm,
the monad construction of the gauge bundle often results 
in a reflexive coherent sheaf 
\nref\dgm{J. Distler, B.R. Greene, and D.R. Morrison,  
``Resolving Singularities
in $(0,2)$ Models,'' Nucl. Phys. {\bf B481} (1996) 289, hep-th/9605222.}%
\dgm\ rather than a bundle. However, at least in the examples studied in
\dgm, reflexive sheaves define non-singular $(0,2)$ CFT's without the exotic
phenomena described above. Given the fact that singularities 
in string theory tend to have a universal local behavior, it is 
reasonable to assume that this is the generic behavior. 

In fact, this is consistent with the spectral cover description 
of the $E_6\rightarrow E_7$ transition. We will show in section 
three that, at the transition point, the $SU(3)$ bundle degenerates 
to a non-reflexive rank three sheaf. Moreover, this sheaf admits a 
natural local decomposition as a  sum of a rank two reflexive sheaf 
and the ideal sheaf of a point. After the transition, the 
heterotic vacuum will be described accordingly as having two 
distinct sectors. We have a perturbative $(0,2)$ CFT part corresponding 
to the reflexive rank two sheaf, which gives an $E_7$ gauge group 
and a certain number of matter multiplets. The second sector consists 
of nonperturbative degrees of freedom localized at certain points 
in the Calabi--Yau threefold, and corresponds to the ideal sheaves. 
This is quite similar to the small instanton effects in six dimensions, 
one of the main differences being the absence of a Coulomb branch. 

This concludes our brief overview of pointlike bundle 
singularities 
in $N=1$ string theories. More details and explicit 
constructions are 
presented in the next sections. We discuss the 
$E_6\rightarrow E_7$ transition
in F-theory, and explain the occurrence of the surfaces $S$ 
in section two. 
The heterotic picture, based on the spectral cover approach, 
is presented in 
section three. Some technical details are postponed to an appendix. 

\newsec{F-theory}
\subsec{Generalities}
Our starting point for the F-theory description is a \CYf\ which is
dual to the heterotic
string on a \CYt\ with a certain $SU(3)\times E_8$ gauge bundle. 
The unbroken $E_6$ gauge group then appears as a singularity in the 
elliptic fiber of the Calabi--Yau fourfold. 

We will choose the heterotic \CYt\ to be elliptically fibered over a base
$B$, which we choose to be the ruled surface $\IF_0 = \IP^1 \times \IP^1$;
this threefold has Hodge numbers $h^{1,1}=3,\; h^{2,1}=243$. The dual \CYf\
$X$ is then 
elliptically fibered over a threefold base $B^{\prime}$ which can be 
viewed as the
total space of the projective bundle ${\bf P}({\cal O}_{B}\oplus
{\cal T})$, where ${\cal T} = {\cal O}_{B}(-\Gamma)$, and $\Gamma$ is 
some
effective divisor in the base $B$, related to the class $\eta$ describing
the heterotic bundle by the relation
$\eta=6c_1 - \Gamma$ 
\nref\FMWb{
R. Friedman, J.W. Morgan, and E. Witten, 
``Vector Bundles and F Theory,'' Commun. Math. Phys. {\bf 187}
(1997) 679, hep-th/9701162.}%
\refs{\FMWb,\toric,\mirror}. Similar models have been considered 
in a different context in \ref\AC{B. Andreas and G. Curio, 
``On discrete Twist and Four-Flux in $N=1$ heterotic/F-theory
compactifications'', hep-th/9908193.}.

The fourfold $\pi': X\rightarrow B^{\prime}$ is described in the vicinity
of a section
$S_0\simeq B$ by the Weierstrass equation
\eqn\threestar{y^2=x^3 + fx + g}
within the bundle ${\cal T}\oplus ({\cal T}^2 \otimes {\cal L}^2) \oplus
({\cal T}^3 
\otimes {\cal L}^3)$. We have ${\cal L}=-K_B$ and ${\cal T} 
\simeq N_{S_0/B^{\prime}}$, the normal
bundle of $B$ in $B^{\prime}$, and we denote by $s$ one of its sections, 
\ie\ $s\in
H^0(B;N_{S_0/B^{\prime}})$. The geometry of the split $IV^*$
singularity 
over the section $S_0\simeq B$ (which corresponds to $E_6$ gauge
group)
\nref\Ber{M.~Bershadsky, K.~Intriligator, S.~Kachru, 
D.R.~Morrison, V.~Sadov, and C.~Vafa, ``Geometric Singularities and 
Enhanced Gauge Symmetries,'' Nucl. Phys. {\bf B481} (1996) 215, 
hep-th/9605200.}%
\Ber\ 
is then encoded in the following expressions for $f$, $g$ and the
discriminant $\delta$:
\eqn\fgdelta{\eqalign{f &= s^3(f_3+sf_4) = s^3 u, \cr
g &= s^4(q_2^2 + sg_5+s^2g_6) = s^4 \widetilde{v}, \cr
\delta &= s^8[4s(f_3+sf_4)^3+27(q_2^2 +
sg_5+s^2g_6)^2]=s^8\delta^{\prime}\cr}} 
where $f_3, f_4, q_2, g_5, g_6$ are sections of certain line bundles over
$B$: $f_3\in H^0(B;{\cal T}\otimes{\cal L}^{\otimes 4})$, 
$f_4\in H^0(B;{\cal L}^{\otimes 4})$,
$q_2\in H^0(B;{\cal T}\otimes{\cal L}^{\otimes 3})$,
$g_5\in H^0(B;{\cal T}\otimes{\cal L}^{\otimes 6})$ and
$g_6\in H^0(B;{\cal L}^{\otimes 6})$.

The fourfold described by \threestar\ can be resolved to a nonsingular
\CYf, in which the singularities of split $IV^*$ type are replaced by
rational curves whose intersection matrix reproduces the Dynkin diagram of
$E_6$ (generically).  No surprises are encountered during this resolution.
One way to describe the resulting F-theory model is as a limit from three
dimensions: first, compactify M-theory on the nonsingular \CYf, then
consider the limit in which all fiber components introduced during the
resolution of the Weierstrass model acquire zero area (leading to enhanced
gauge symmetry), and finally, take
the F-theory limit by sending the area of the elliptic fibers to zero,
opening up a new effective dimension.

The transition (un-Higgsing) to unbroken $E_7$ gauge group is described
in F-theory terms by the condition $q_2=0$, which results in the singularity
being enhanced to $III^*$ fibers. Let us describe in some detail the 
geometry of this singular fourfold. 

Under the condition $q_2=0$, the discriminant $\delta$ actually has a factor
of $s^9$:
\eqn\newdelta{
\delta = s^9[4(f_3+sf_4)^3+27s(
g_5+sg_6)^2]=s^9\delta^{\prime\prime}
}
leading to $III^*$ fibers.
The component $\Delta^{\prime\prime}$ of the discriminant is
defined by the equation $\delta^{\prime\prime}=0$; this latter equation
is a cubic equation 
in $s$ whose  discriminant with respect to $s$ is given by:
\eqn\noua{
{\rm discrim}(\delta^{\prime\prime}) = 2^4\cdot3^9\cdot
(f_4g_5-f_3g_6)^3(f_3^2f_4g_5+g_5^3-f_3^3g_6)} 
Let $\Sigma$ denote the locus $f_3=0$, therefore the matter curve is 
$\Sigma =4c_1 - \Gamma$. Then, it is easy to see that the
locus $g_5=0$ is
precisely $\eta=6c_1-\Gamma$. Finally, by $\xi$ and $\omega$ we denote 
the loci $f_4g_5-f_3g_6=0$ and $f_3^2f_4g_5+g_5^3-f_3^3g_6=0$ respectively. 
The singularity type is enhanced to $II^{*}$ over the 
matter curve $\Sigma$, which is the intersection locus of $S_0$ with
the nodal part of the discriminant, $D_1$. More interesting, the
vanishing orders jump to $(4, 6, 12)$ over the intersection locus 
$\eta\cap\Sigma$.  (In the familiar case of an elliptic surface, this would
be the signal that the Weierstrass model was not
minimal.  However, in the present context there is no birational change
which can be made which would reduce those orders of vanishing.)
 The set where the vanishing orders jump to $(4, 6, 12)$ is precisely the
singularity set of the corresponding  heterotic 
sheaf (which would otherwise be a bundle, were it not for the presence of
this locus). There is a cusp curve $\Xi$ inside $D_1$, which projects onto
the curve $\xi$ in $B$. The geometry of the singular 
fourfold is sketched in Figure 1. 

\ifig\sing{Geometry of the singular
fourfold.}{\epsfxsize3.5in\epsfbox{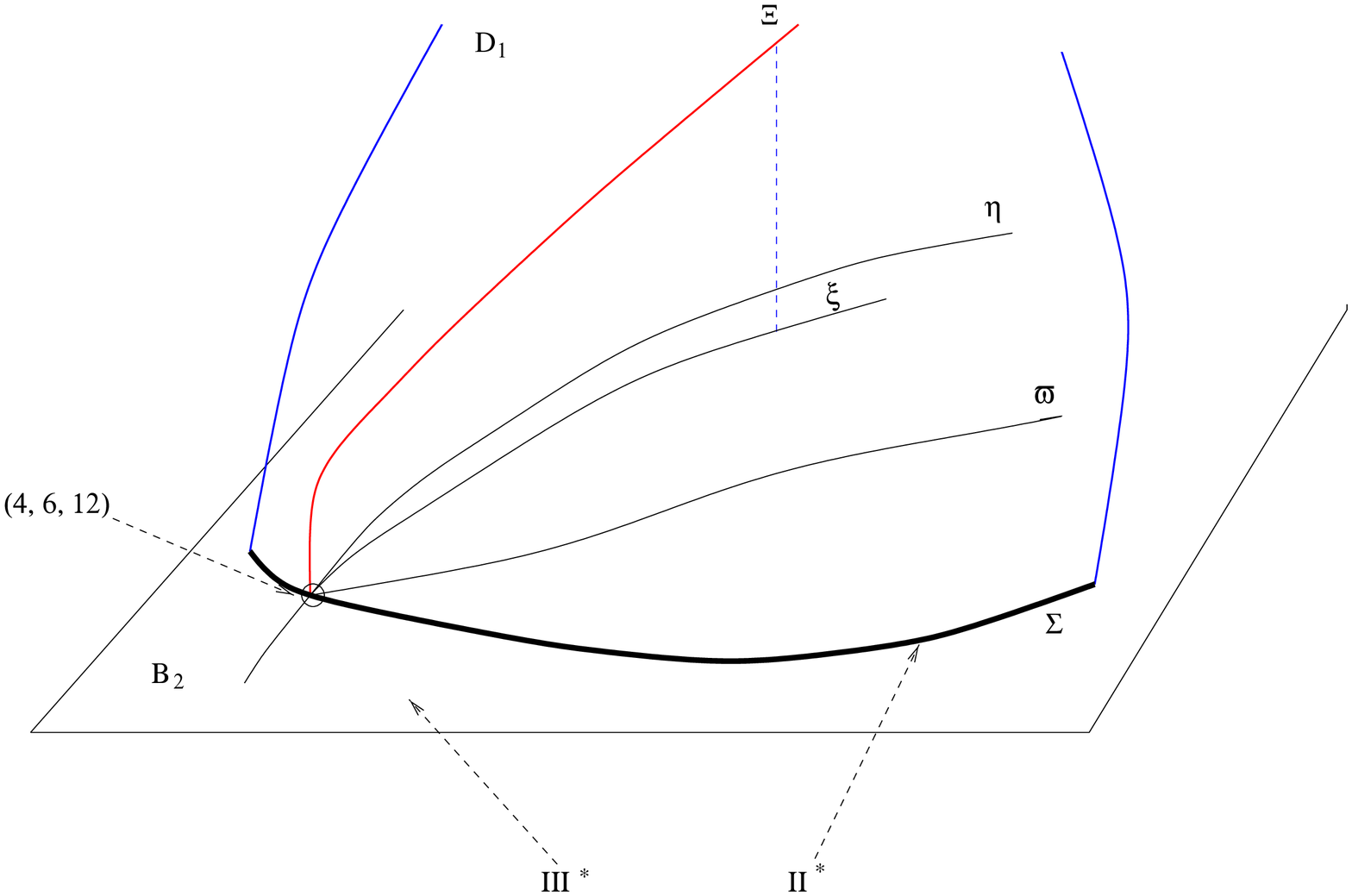}}

As explained in the introduction, resolution
of this $III^*$ locus results in the appearance of entire {\sl complex
surfaces}\/ over the locus $\eta\cap\Sigma$ in $B^{\prime}$. This 
phenomenon is 
most efficiently observed by performing a weighted blowup of the 
Weierstrass model, which we now proceed to describe.

\subsec{The Weighted Blowup}
The weighted blowup is performed 
by introducing an additional variable
$\lambda$, and assigning weights as follows:
$$\eqalign{
\lambda\hbox{\hskip10pt} &s\hbox{\hskip10pt} x\hbox{\hskip10pt} y\cr 
-1\hbox{\hskip10pt} &1\hbox{\hskip10pt} 2\hbox{\hskip10pt} 3\cr}$$ 
We now rewrite \threestar\ as a homogeneous degree 5 equation in the variables
$\lambda,s,x,y$ as follows
\eqn\wbup{y^2\lambda= x^3\lambda + s^3 u x + s^5 v}
This is the weighted blowup of \threestar. In the patch $\lambda\neq 0$,
it is equivalent to \threestar, but when $\lambda=0$, we get
\eqn\surface{s^3(ux+vs^2)=0}
Note that over the point $u=v=0$, this vanishes identically. Thus
$u,v\in \IC^2$ parametrise a family of hypersurfaces in $\IP^{(1,2,3)}$ except
at $u=v=0$, when we obtain all of $\IP^{(1,2,3)}$. This is precisely the
complex surface mentioned before.  Its occurance is a direct consequence of
the $(4, 6, 12)$ vanishing orders of the discriminant along $u=v=0$.

\lref\MIR{R.~Miranda, ``Smooth models for elliptic
threefolds,'' in {\it The
Birational Geometry of Degenerations}, R. Friedman and D.R. Morrison 
eds, Birkh\"auser, 1983, p.~85.}%

The weighted blowup is really only the first stage of a complete toric
resolution of the $III^*$ singularity, and thus does not introduce any extra
K\"ahler classes beyond those already needed for the usual resolution of
singularities. 
We present some further evidence for this lack of additional K\"ahler
classes by studying an explicit toric 
example in the next subsection.\foot{It is also possible to perform
an explicit local resolution using the technique of 
\MIR, which would be related by generalized flops to the resolution 
presented in this section.}

\subsec{Toric Example}

We now construct an explicit toric model of a \CYf\ elliptically 
fibered over
base $B_3=\IF_{056}$ (\ie\ $\Gamma = 5C_0 + 6f$, where $C_0$ and $f$ 
are the
classes in $B_2=\IF_0$), with a section $S_0\simeq B_2 = \IF_0$ of split
$IV^*$ singularities. This fourfold is dual to heterotic strings 
compactified
on the $(3,243)$ \CYt\ which is elliptically fibered over $B_2= \IF_0$, 
with
an $SU(3)\times E_8$ bundle with $\eta_{SU(3)}= 6c_1(B_2)-\Gamma =
7C_0 + 6f$,
and $\eta_{E_8}= 6c_1(B_2)+\Gamma = 17C_0 + 18f$.

From index theorems and anomaly cancellation (see, for
instance \refs{\toric,\mirror}), we expect the following Hodge numbers,
$h^{1,1}=10,\; h^{3,1}-h^{2,1}=9231,\; \chi = 55494$.

The \CYf\ may be constructed as a hypersurface in a toric variety 
following
the prescription of \refs{\toric,\mirror}.
The dual polyhedron $\nabla$, which
encodes the divisors of the polyhedron has vertices:
$$\eqalign{&{\tt (-1, 0, -6, 2, 3),\; (0, -1, -5, 2, 3),\; 
(0, 0, 0, -1, 0),\;
(0, 0, 0, 0, -1), \; (0, 1, 0, 2, 3),}\cr
&{\tt (0, 0, 1, 2, 3),\; (1, 0, 0, 2, 3),\; (0, 0, -2, 1, 1),\;
(0, 0, -1, 0, 0)}\cr}$$
Standard toric methods \refs{\toric,\mirror} give, for the Hodge numbers
of the fourfold, $h^{1,1} = 10,\; h^{3,1} = 9231,\; h^{2,1} = 0,\; 
h^{2,2} = 37008,\; \chi = 55494$, in agreement with expectations. 
Note that
the Euler characteristic is not divisible by $24$, indicating that the 
model
has a background $G$-flux turned on. Moreover, it
is possible to find a triangulation of the polyhedron consistent with 
its
elliptic fibration structure, such that each of the top dimensional cones
has unit volume, guaranteeing smoothness of the corresponding
\CYf. We assert that this
polyhedron gives the F-theory dual of the heterotic vacuum described 
above.

We can now study the effect of un-Higgsing the unbroken $E_6$ gauge 
group to
$E_7$. The heterotic bundle is now $SU(2)\times E_8$, with
$\eta_{SU(2)}=6c_1(B_2)-\Gamma = 7C_0 + 6f$, and $\eta_{E_8}$ unchanged. 
Index
theorems and anomaly cancellation predict, for the dual \CYf, 
Hodge numbers
$h^{1,1}=11,\; h^{3,1}-h^{2,1}=9221,\; \chi = 55440$.

The dual polyhedron $\nabla$ describing the \CYf\ has vertices
$$\eqalign{&{\tt (-1, 0, -6, 2, 3),\; (0, -1, -5, 2, 3),\; 
(0, 0, 0, -1, 0),\;
(0, 0, 0, 0, -1),\; (0, 1, 0, 2, 3),}\cr
&{\tt (0, 0,-1, 0, 1), \; (0, 0, 1, 2, 3), \; (1, 0, 0, 2, 3)}\cr}$$ 
The fourfold has the following Hodge numbers, in accordance with our
expectations: $h^{1,1} = 11,\; h^{3,1} = 9221, \; h^{2,1} = 0, \;
h^{2,2} = 36972, \; \chi = 55440$. Once again, it is possible to find a
triangulation of the of polyhedron consistent with its elliptic fibration
structure, such that each of the top dimensional cones has unit volume,
guaranteeing smoothness of the corresponding \CYf.

It should be emphasized here that no extra K\"ahler classes other than the
ones corresponding to the resolution of the $III^*$ locus are present 
in the
fourfold. Since we expect, on general grounds, that the resolution of the
singularity yields an entire complex surface over specific points 
in $B_3$, we
conclude that {\sl the appearance of the complex surface in the 
resolution of
the $III^*$ locus does not introduce any extra K\"ahler classes}.

The corresponding F-theory model has some features whose physical effects
are difficult to explain in detail.  We begin as before in three dimensions,
with M-theory compactified on the nonsingular \CYf.  When we allow the
rational curves in the fibers to shrink to zero area, again we get enhanced
gauge symmetry, but this time there are surfaces shrinking to points as
well as curves shrinking to points.  Wrapping the M-theory fivebrane on
such surfaces suggests that the spectrum should contain light strings,
while wrapping the M-theory membrane on curves within such surfaces would
produce a tower of light particle states.
All of these states are presumably present as well in the F-theory limit.

\subsec{Comparison with Codimension-Two}

It is worthwhile making a comparison between the geometry of these
codimension-three singularities, and the analogous phenomenon in
codimension-two.  In the latter case, the F-theory interpretation of
a small instanton singularity is that the total space of the elliptic
fibration has
acquired a singularity which can be resolved by a combination of blowing up
the base of the fibration and blowing up the total space \mvtwo.  (It is the
blowup of the base which leads to an additional branch of the moduli
space.)   In the codimension-three case, however, blowing up the base is
not possible, because it destroys the Calabi--Yau condition.

To see this, consider first a simple model of the codimension-two
phenomenon, represented by the Weierstrass equation
\eqn\codimtwo{
y^2=x^3+s^4x+s^5t.
}
One of the coordinate charts when blowing up the base is $s_1=s$,
$t_1=t/s$; in that chart, the Weierstrass equation becomes
\eqn\codimtwobu{
y^2=x^3+s_1^4x+s_1^6t_1.
}
To make this new Weierstrass equation minimal, we must also change
coordinates in $x$ and $y$, using $x_1=x/s^2$, $y_1=y/s^3$.  Our final
Weierstrass equation is then
\eqn\codimtwomin{
y_1^2=x_1^3+x_1+t_1.
}
(This two-step change of variables corresponds to the two-step geometric
process of a blowup and a flop which was  used in \mvtwo\ to describe this
transition.) 

According to the Poincar\'e residue construction, the holomorphic
three-form was originally represented by 
\eqn\holotwo{
{{dx \wedge ds \wedge dt} \over 2y},
}
where $2y$ represents the partial derivative of \codimtwo\ with respect to
the variable $y$ (which is not present in the numerator).
In the new coordinate system (the minimal model of the blowup) this
becomes 
\eqn\holotwomin{
{{(s_1^2 dx_1) \wedge ds_1 \wedge (s_1 dt_1)}\over 2 s_1^3 y_1}
={{dx_1 \wedge ds_1 \wedge dt_1} \over 2y_1}.
}
Since this latter is the Poincar\'e residue representation of a holomorphic
three-form for the blown up threefold \codimtwomin, our original three-form
has acquired 
neither a zero nor a pole during this process.  Thus, both threefolds can
be Calabi--Yau and there is a transition between them.
(Note that even though we only made the computation in a single coordinate
chart, the order of zero or pole of the holomorphic three-form would be the
same in {\it any}\/ coordinate chart, so this is actually a complete
argument.)

By contrast, let us make a similar computation for a fourfold, starting from
the Weierstrass equation
\eqn\codimthree{
y^2=x^3+s^3ux+s^5v.
}
We can represent one of the coordinate charts of the blowup by $s_1=s$,
$u_1=u/s$, $v_1=v/s$; in that chart, the Weierstrass equation becomes
\eqn\codimthreebu{
y^2=x^3+s_1^4u_1x+s_1^6v_1.
}
We again get a non-minimal Weierstrass model, which can be made minimal by
the further coordinate change $x_1=x/s^2$, $y_1=y/s^3$.  Our final
Weierstrass equation is then
\eqn\codimthreemin{
y_1^2=x_1^3+u_1x_1+v_1.
}

The holomorphic four-form was originally represented by
\eqn\holothree{
{dx \wedge ds \wedge du \wedge dv} \over 2y,
}
and in the minimal model of the blowup this becomes
\eqn\holothreemin{
{{(s_1^2 dx_1) \wedge ds_1 \wedge (s_1 du_1) \wedge (s_1 dv_1)}\over 2
s_1^3 y_1} ={{s_1 dx_1 \wedge ds_1 \wedge du_1 \wedge dv_1} \over 2y_1}.
}
Since this is $s_1$ times the Poincar\'e residue representation of a
holomorphic 
four-form for the blown up fourfold \codimthreemin, our original four-form
has acquired a zero along the exceptional divisor $s_1=0$.  Thus, at most
one of these two fourfolds can be Calabi--Yau (\ie\ have a {\it
non-vanishing}\/ holomorphic four-form), and there is no physical transition
between them.

\newsec{Singular Bundles and Transitions}

\subsec{Spectral Data}

We begin with a short review of the spectral cover approach to bundles
on elliptic fibrations \FMW. Let $\pi:Z\rightarrow B$ be a smooth 
elliptic Calabi--Yau variety 
with a section $\sigma:B\rightarrow Z$. As usual, we assume that 
$Z$ is moreover a cubic hypersurface in 
$\IP\left(\co_B\oplus\cl^2\oplus\cl^3\right)$, where $\cl=\co_B(-K_B)$
is the anticanonical line bundle of the base. 

According to \FMW,\ the moduli space of rank $n$ semistable bundles 
with trivial determinant on a smooth elliptic curve $E$ is isomorphic 
to the linear system $|np_0|\simeq \IP^{n-1}$, where $p_0$ is the 
origin of $E$. This construction works for families of elliptic curves 
as well, if the singular fibers are either nodal or cuspidal curves. 
For the Weierstrass model $Z\rightarrow B$ introduced before, this 
yields a relative coarse moduli space $\cp_{n-1}\rightarrow B$ 
which is isomorphic to the relative projective space 
$\IP\left(\co_N\oplus\cl^{-2}\oplus\cdots\oplus\cl^{-n}\right)$.
If $V\rightarrow Z$ is a rank $n$ bundle whose restriction to 
every fiber $E_b$ is semistable and regular with trivial determinant, 
then $V$ determines a section $A:B\rightarrow \cp_{n-1}$.
Such a section $A$ is uniquely given by a line bundle $\cm$ over $B$, 
and sections $a_i\in H^0\left(B,\cm\otimes\cl^{-i}\right)$, 
$i=0,2,\ldots,n$. 

The converse is not true, \ie\ a section $A$ does not uniquely 
determine a bundle $V$. Friedman, Morgan, and Witten  \FMW\
construct certain basic bundles $V_{A,a}$ associated to a section $A$ 
together with an integer $a\in \IZ$. The construction is rather involved 
and it will not be reviewed here in detail. After some work, it can be
shown that it is equivalent to the standard spectral cover construction 
\FMW. Namely, the section $A$ determines a spectral cover 
$C_A\subset Z$ which belongs to the linear system 
$|n\sigma+\pi^*\alpha|$, where $\alpha=c_1(\cm)$. In order to construct 
$V_{A,a}$, let us consider the following diagram \FMW
$$\xymatrix{
  T_A    \ar[r]^{\nu_A} \ar[d]_{\rho_A} &  Z \ar[d]^{\pi}\\
C_A  \ar[r]^{g_A} & B.\\
}$$
Note that $T_A\rightarrow C_A$ is an elliptic fibration with a 
section $\Sigma_A=\nu_A^*\sigma$. There are certain 
natural Weil divisor classes  on $T_A$: the diagonal $\Delta$ 
obtained by restriction from $Z\times_B Z$ and the class 
$\rho_A^*(F)$, where $F=C_A\cdot\sigma$.\foot{$\Delta$ is not a 
Cartier divisor near the singular locus of $T_A$.}
Then we have 
\eqn\bundleA{
V_{A,a}=(\nu_A)_*\co_{T_A}\left(\Delta-\Sigma_A-a\rho_A^*F\right).}

At this point, it may be helpful to compare the present approach to 
the more familiar construction of \FMWb. Let $\cp\rightarrow Z\times_B Z$ 
denote the universal Poincar\'e line bundle and let $\cn$ denote a 
line bundle over the spectral cover $C_A$. Note that 
\eqn\poincare{
\cp=\co_{Z\times_BZ}\left(\Delta-\sigma_1-\sigma_2\right)\otimes 
\cl^{-1}.}
Then we construct a bundle $V$ associated to the pair $(C_A,\cn)$
\eqn\bundleB{
V=\nu_{A*}\left(\rho_A^*\cn\otimes\cp_{T}\right).}
We can establish a direct relation between \bundleA\ and \bundleB\ by 
noting that $\sigma_1|_{T_A}=\Sigma_A$ and $\sigma_2|_{T_A}=F$. 
Therefore we 
must have 
\eqn\rel{
\cp|_{T_A}=\co_{T_A}(\Delta - \Sigma_A - F)\otimes \cl^{-1},\qquad
\cn=\co_{C_A}(-(a-1)F)\otimes\cl.}
We can construct more general bundles by twisting by an arbitrary 
line bundle $\cs$ pulled back from $B$.

The Chern classes of the bundles $V_{A,a}$ are given by \FMW
\eqn\chern {
\ch\left(V_{A,a}\right)=e^{-\alpha}{1-e^{(a+n)L}\over 1-e^L}-
{1-e^{aL}\over 1-e^L}+e^{-\sigma}(1-e^{-\alpha}),}
where $L=c_1(\cl)$.
From now on, we will supress the indices $A,a$ for simplicity. The 
bundles will be denoted by $V_n$ in order to emphasize their rank. 

\subsec{Reducible Spectral Cover}

Generically, the spectral cover $C$ is smooth and irreducible, 
at least if the line bundle $\cm$ is sufficiently ample on $B$. 
For physics reasons, we have nevertheless to understand the bundles 
$V$ associated to reducible spectral covers. Following \FMW\ 
(sect 5.7) we consider here the case when the spectral cover $C$ 
splits as a sum $C^\prime + \sigma$, but is otherwise generic. 
As explained in detail in \FMW, such a degeneration of $C$ corresponds 
to a section $A:B\rightarrow \cp_{n-1}$ which lies in the hyperplane 
$\cp_{n-2}\subset \cp_{n-1}$. $C^\prime$ corresponds to a rank 
$n-1$ bundle $V_{n-1}$ and we have the following elementary 
modification
\eqn\elmodifA{
0\rightarrow V_n\rightarrow V_{n-1}\oplus \pi^*\cl^a
\rightarrow \pi^*\cl^a|_{\pi^*F}\rightarrow 0.}
Here we view $F=C\cdot\sigma$ as a divisor on $\sigma\simeq B$, 
hence $\pi^*F$ is the vertical divisor above $F$. In terms of 
divisor classes on $B$, 
\eqn\inters{
F=\alpha-(n-1)L.}
Note that by construction, there is a surjective 
morphism $V_n\ra V_{n-1}$ whose kernel is isomorphic to 
$\pi^*\cl^a(-F)$. Therefore we have an exact 
sequence 
\eqn\exseqA{
0\rightarrow \pi^*\cl^{a}(-F)\rightarrow V_n\rightarrow 
V_{n-1}\rightarrow 0.}
The rank changing transition described above 
is valid for $n$ sufficiently large 
such that $A$ is a regular section of both $\cp_{n-2}$ 
and $\cp_{n-1}$. If $n$ is small, \ie\ $n\leq 3$, this is no longer 
true and the above picture must change. Understanding the new picture will
be our main  
goal in the remaining part of this section.

\subsec{Singular Bundles and Reflexive Sheaves}

For small values of  $n$ (such as $n=3$ in the present case)
we cannot have regular sections $A:B\ra \cp_{n-2}$.
Recall that a section $A:B\ra \cp_1$ 
would be defined by a line bundle $\cm$ on $B$ together with two 
sections $a_0\in H^0(B,\cm)$, 
$a_2\in H^0(B,\cm\otimes \cl^{-2})$. 
Since the base $B$ is two dimensional, the sections $a_0,a_2$ will 
have generically common zeroes at finitely many points ${b_1,\ldots,b_k}$ 
on $B$. This means that the section $A$ is defined only on the complement 
of $\{b_i\}$ in $B$. The closure of the image of $A$ in $\cp_1$ is 
isomorphic to the blowup ${\widetilde B}$ of $B$ at the points 
$\{b_i\}$.
The exceptional curves on ${\widetilde B}$ coincide with the $\IP^1$ 
fibers projecting to the points $B_i$. We will call such a section $A$
a quasisection. 

Given a quasisection $A$, and an integer $a\in \IZ$, we can construct 
a rank $2$ bundle $V_2$ over the open subset $Z\setminus \cup E_i$, 
where $E_i=\pi^{-1}(b_i)$.
However, it is not clear if $V_2$ can be extended as a bundle over
the threefold $Z$. In general, even if an extension exists, we expect 
some kind of singular behavior along the elliptic fibers $E_i$. 

This problem has been addressed in \FMW\ where it has been shown that 
$V_2$ can always be extended as a sheaf, and the singular behavior 
depends on the parity of $a$. We now recall the main construction and 
the central results. The idea is, roughly, to work over the blown up base
$\wb$ where we can apply the standard construction and then push-forward 
to $B$. Let us consider the diagram 
$$\xymatrix{
  {\widetilde \cp_{1}}    \ar[r] \ar[d] &  \cp_{1} \ar[d] \\
{\widetilde B} \ar[r] & B.\\
}$$
The pullback of the rational section $A$ to ${\widetilde \cp}_{1}$ 
is  a regular section ${\widetilde A}:{\widetilde B}\rightarrow 
{\widetilde \cp}_{1}$. Therefore we can construct a rank 2 bundle 
${\widetilde V}_2$ over ${\widetilde Z}={\widetilde B}\times_B Z$.
Note that $\wz$ is the blowup of $Z$ along the 
elliptic curves $E_i$, with exceptional divisors $D_i=E_i\times \IP^1$.
Let $q:\wz\rightarrow Z$ denote the blowup map. 
Then, the direct image $q_*\wv_2$ defines in principle 
an extension of $V_2$. However, in order to keep the singular behavior 
along the elliptic fibers $E_i$ under control, we have to perform a 
more elaborate construction. 
The result depends on the behavior of $\wv_2$ along the exceptional 
divisors of $D_i$, which in turn depends on the parity of $a$. 

If $a$ is even, the bundle $V_2$ can be extended over $Z$ as a bundle 
whose restriction to the fibers $E_i$ is unstable. This is a rather 
mild singular behavior and we will not discuss this case further in 
this paper. 

If $a$ is odd, the bundle $V_2$ can be extended as a reflexive sheaf 
over $Z$, whose construction will be reviewed here in some detail. 
Note that we will denote the extension also by $V_2$, the meaning being 
clear from the context. In order to avoid unnecessary complications, 
we will also fix $a=1$ and work near a fixed fiber $E_i$, dropping 
the index $i$. 

The restriction of $\wv_2$ to the exceptional divisor $D=E\times \IP^1$ 
is \FMW\
\eqn\jump{
\wv_2|_{{\IP^1}\times p}=\left\{\matrix{
& \co_{\IP^1}(-1)\oplus \co_{\IP^1}(-1),
& \qquad\hbox{if}\ p\neq p_0\cr
& \co_{\IP^1}\oplus\co_{\IP^1}(-2),&\qquad \hbox{if}\
p=p_0.\cr}\right.}
where $p_0$ is the origin of $E$. 
In order for the direct image to be well behaved,  we would like 
$\wv_2$ to be trivial along the exceptional fibers $\IP^1$. 
This can be realized by first twisting by $\co_{\wz}(-D)$ and then 
performing an elementary modification along the curve $\IP^1\times p_0$.
However, an elementary modification in codimension two does not result 
in a locally free sheaf. Therefore, we need to perform a second blowup 
along ${\IP^1}\times p_0$. Let $g:Z_1\rightarrow \wz$ denote the blowup map
with exceptional divisor $D_1\simeq\IP^1\left(\co_{\IP^1}\oplus 
\co_{\IP^1}(-1)\right)$. Let $D^\prime$ denote the proper transform 
of $D$ to $Z_1$, that is, $g^*D=D^\prime + D_1$; $D^\prime$ and $D_1$ 
intersect transversely along the negative section of $D_1$. Note that 
$D^\prime$ is a rationally ruled surface over $E$. Finally we can 
contract $D^\prime$ along its fibers obtaining a threefold $Z_2$ 
which is the blowup of $Z$ at the point $p_0\in E$. The exceptional 
divisor $D_2\simeq \IP^2$ is obtained by contracting the 
negative section on $D_1\simeq \IF_1$. This sequence of birational 
transformations can be summarized in the following diagram 
$$\xymatrix{
  Z_1    \ar[r]^{g} \ar[d]_{p} &  \wz \ar[d]^{q} \\
 Z_2  \ar[r]^{f} & Z.\\
}$$

Now let us describe the bundle construction \FMW.
We first pull back 
$\wv_2(-D)$ to $Z_1$ and perform an elementary modification 
along $D_1$
\eqn\elmodifB{
0\rightarrow V^\prime_2\rightarrow g^*\wv_2(-D)\rightarrow 
j_*\left(\co_{D_1}(-f)\right)\rightarrow 0,}
where $j:D_1\rightarrow Z_1$ is the inclusion and $f$ is the 
fiber class of $D_1\simeq \IF_1$. It turns out that $V_2^\prime$
is locally free and uniformly trivial along the fibers of $D^\prime$. 
Then, it follows that $p_*V_2^\prime$ is also locally free, and with 
some work it can be shown that its restriction to the exceptional 
divisor $D_2=\IP^2$ is isomorphic to $T_{\IP^2}(-1)$. 

The final step is to take the direct image $V_2=f_*p_*V_2^\prime$ on 
$Z$. Given the above behavior along $D_2$, it can be shown using the 
formal functions theorem \FMW\ that $V_2$ is a reflexive sheaf on $Z$. 
This concludes the extension of the rank two bundle defined over 
$Z\setminus \cup E_i$ to $Z$. 

\subsec{The Transition}

In this subsection we generalize the rank changing transition 
discussed in section 3.2 to $n=3$. In order to fix ideas, recall 
that we start with a smooth rank three bundle $V_3$ corresponding to 
a regular section of $\cp_2\ra B$. Our goal is to understand 
the bundle degeneration associated to a deformation of this 
regular section to a quasisection $A:B\ra \cp_1\subset \cp_2$.
For simplicity, we will assume that there is a single point $b\in B$ 
where $A$ fails to be a section. The generalization to a finite 
collection of points $\{b_1,\ldots,b_k\}$ is straightforward. 

The main idea is to work again on the blown up space $\wz$ where 
we can apply the standard theory, and then push forward to $Z$. 
However, there are extra complications arising from the necessity of 
a second blowup, as discussed above.
The quasisection $A$ pulls back to a section ${\widetilde A}$ of 
${\widetilde \cp}_2$ over $\wb$ whose image lies in ${\widetilde \cp}_1$, 
but is otherwise generic. Therefore we can construct a rank three 
bundle as in section 3.2 
\eqn\elmodifC{
0\rightarrow \wv_3\rightarrow \wv_2\oplus \cl^a\rightarrow 
\cl^a|_{\wf}\rightarrow 0,}
where 
\eqn\relation{
\wf=\wal - 2L=\alpha - D -2L.} 
Note that $\wf, \alpha, L$ should be understood now as divisor classes
on $\wz$ obtained by pull-back from $\wb$. We have dropped the 
explicit notation ${\widetilde \pi}^*$ for simplicity. 
In the following, it will be more convenient to use the alternative 
presentation 
of $\wv_3$
\eqn\exseqB{
0\ra \cl^a(-\wf) \rightarrow \wv_3\rightarrow \wv_2
\rightarrow 0.}
Next we proceed step by step, by analogy with the construction 
of the reflexive sheaf $V_2$ in the previous section.
Twisting by $\co_{\wz}(-D)$, and taking into account \relation,\
we obtain the following exact sequence 
\eqn\elmodifE{
0\rightarrow \cl^{a+2}(-\alpha)\rightarrow \wv_3(-D)\ra
\wv_2(-D) \ra 0.}
We have 
\eqn\restrG{\eqalign{
& \wv_2(-D)|_{\IP^1\times p_0}=\co_{\IP^1}(1)\oplus\co_{\IP^1}(-1)\cr
& \wv_3(-D)|_{\IP^1\times p_0}=\co_{\IP^1}(1)\oplus\co_{\IP^1}(-1)
\oplus \co_{\IP^1}\cr}}
since \elmodifE\ splits over ${\IP^1}\times p_0$. Moreover, since 
$ \cl^{a+2}(-\alpha)$ is trivial along the fibers of $D\ra E$, 
it follows from \jump\ and \elmodifE\ that 
\eqn\genrestr{
\wv_3(-D)|_{\IP^1\times p}=\co_{\IP^1}\oplus\co_{\IP^1}\oplus\co_{\IP^1}}
for $p\neq p_0$.
Following the steps outlined above, we pull back \elmodifE\ 
to $Z_1$, and perform an elementary modification along $D_1$. 
Note that \restrG\ implies
\eqn\restrF{\eqalign{
&g^*\wv_2(-D)|_{D_1}=\co_{D_1}(f)\oplus\co_{D_1}(-f)\cr
&g^*\wv_3(-D)|_{D_1}=\co_{D_1}(f)\oplus\co_{D_1}(-f)\oplus \co_{D_1},\cr}}
therefore we can perform elementary modifications of $g^*\wv_2(-D),
g^*\wv_3(-D)$ along $D_1$. We obtain the following commutative 
exact diagram
$$\xymatrix{
           & 0 \ar[d]                      & 0 \ar[d]                     & 0 \ar[d]                       &  \\
 0  \ar[r] &\cl^{a+2}(-\alpha)      \ar[r] \ar[d] &V_3^\prime      \ar[r] \ar[d] &V_2^\prime\ar[r] \ar[d] &0  \\
 0  \ar[r] &\cl^{a+2}(-\alpha)    \ar[r] \ar[d] &g^*\wv_3(-D)    \ar[r] \ar[d] &g^*\wv_2(-D)\ar[r] \ar[d] &0\\
 0  \ar[r] & 0 \ar[r] \ar[d] &j_*\co_{D_1}(-f)\ar[r] \ar[d] & j_*\co_{D_1}(-f) \ar[r] \ar[d] &0  \\
           & 0                             & 0                            & 0                              & \\
}$$
Given the above diagram, the results of \FMW\ show that 
$f_*p_*V_3^\prime$ is a rank 3 reflexive sheaf of $Z$. 
One might think that this is the solution to our problem, namely that 
$f_*p_*V_3^\prime$ can then be deformed to a smooth rank 3 bundle 
corresponding to a generic section of $\cp_2\ra B$. It turns out 
however that this is not true since the sheaf $f_*p_*V_3^\prime$
has the {\it wrong topological invariants}. This fairly elaborate 
computation is postponed to appendix A.

In order to obtain a sheaf with the correct topological invariants, 
we have to perform a further elementary modification on $\vp_3$. 
By restricting the top row of the above diagram 
to $D_1$, we obtain the following exact sequence 
\eqn\exseqN{
0\ra \co_{D_1}\ra \vp_3|_{D_1}\ra \vp_2|_{D_1}\ra 0,}
where $\vp_2|_{D_1}\simeq p^*T_{\IP^2}(-1)$, \FMW\ (claim 2, pg 87).
It can be checked that $\hbox{Ext}^1(p^*T_{\IP^2}(-1),\co_{D_1})=0$, 
therefore 
$\vp_3|_{D_1}\simeq \co_{D_1}\oplus p^*T_{\IP^2}(-1)$. 
Now we can perform the elementary modification 
\eqn\elmodifF{
0\ra\vpp_3\ra\vp_3\ra j_*\co_{D_1}\ra 0.}
We claim that $f_*p_*\vpp_3$ represents the solution to our 
problem. 

In the following we will examine the properties of $f_*p_*\vpp_3$
and derive a simple relation between 
$f_*p_*\vpp_3$ and $f_*p_*\vp_2$. 
Along the way we will prove that all the higher direct images of 
the form $R^ip_*\vpp_3$, $R^if_*(p_*\vpp_3)$ vanish. Combined with the 
Chern class computations of appendix A, this shows that 
$f_*p_*\vpp_3$ has the right topological invariants. 

Recall that the map $p:Z_1\ra Z_2$ contracts the fibers of $D^\prime$, 
which is a $\IP^1$-fibration over $E$. Note that  $\vpp_3$ 
is trivial along all fibers except possibly the fiber projecting to 
$p_0\in E$. This follows from \genrestr.\ The fiber over $p_0$ coincides 
with the negative section $l$ of $D_1$. By construction, we have an exact 
sequence 
\eqn\exseqO{
0\ra \co_{D_1}(l+f)\ra \vpp_3|_{D_1}\ra p^*T_{\IP^2}(-1)\ra 0}
which restricts to 
\eqn\exseqOA{
0\ra \co_{\IP^1}\ra \vpp_3|_l\ra \co_{\IP^1}\oplus \co_{\IP^1}\ra 0.}
This shows that $\vpp_3$ is also trivial along $l$. 
Therefore, we can contract the fibers of $D^\prime$, obtaining a vector 
bundle $p_*\vpp_3$ on $Z_2$. Moreover, using the base change theorem, 
it can be shown that 
$R^ip_*\vpp_3=0$ for all $i\geq 1$, as claimed before. 
Then, the exact sequence \elmodifF\ induces a similar exact 
sequence on $Z_2$
\eqn\exseqP{
0\ra p_*\vpp_3\ra p_*\vp_3\ra j_*\co_{D_2}\ra 0}
which shows that $p_*\vpp_3$ is an elementary modification of 
$p_*\vp_3$. By computing extensions, it can be shown that 
\exseqO\ must be split, $\vpp_3|_{D_1}=\co_{D_1}(l+f)\oplus 
p^*T_{\IP^2}(-1)$. After contraction, this yields
$p_*\vpp_3|_{D_2}=\co_{\ip^2}(1)\oplus T_{\IP^2}(-1)$, which implies
$R^if_*(p_*\vpp_3)=0$ 
for all $i\geq 1$. Therefore we have an exact sequence 
\eqn\exseqQ{
0\ra f_*p_*\vpp_3\ra f_*p_*\vp_3\ra {\cal O}_{p_0}\ra 0.}
In order to derive a direct relation with $f_*p_*\vp_2$, note that 
we have another short exact sequence on $Z$
\eqn\exseqR{
0\ra \cl^{a+2}(-\alpha)\ra f_*p_*\vp_3\ra f_*p_*\vp_2\ra 0.}
Combining \exseqQ\ and \exseqR\ we obtain 
\eqn\exseqS{
0\ra {\cal J}_{p_0}\otimes \cl^{a+2}(-\alpha)
\ra f_*p_*\vpp_3\ra f_*p_*\vp_2\ra 0,}
which is the promised relation. 

Moreover, \exseqQ\ shows that the local behavior of the 
sheaf $f_*p_*\vpp_3$ near $p_0$ is of the form ${V_2}_{p_0}
\oplus {\cal J}_{p_0}$ where ${V_2}_{p_0}$ is a reflexive rank two 
$\co_{Z,p_0}$-module. The generalization of \exseqS\ to a 
finite collection of singular points $\{b_i\}$ is 
straightforward---we simply replace ${\cal J}_{p_0}$ by ${\cal
J}_{p_{01}+\ldots + 
p_{0k}}$.
\subsec{The Heterotic Transition}

The bundles $V_{A,a}$ studied above are perhaps the most natural 
in the spectral cover construction, but they are not quite 
suitable for physics. In order to define a consistent heterotic 
background we need bundles with $c_1(V)=0$ and $c_2(V)=c_2(X)$. 
The second condition can be relaxed if we allow heterotic 
fivebranes wrapping holomorphic curves in the Calabi--Yau 
threefold. In order to satisfy the first condition, one can try to 
tensor the bundles $V_{A,a}$ by a line bundle $\cn_0$ pulled back 
from $B$. By expanding \chern,\ we find 
\eqn\det{
c_1\left(V_{A,a}\otimes \cn_0\right)= 
\left(an+{n(n-1)\over 2}\right)L-(a+n-1)\alpha+nc_1(\cn_0).}
Notice that if $n$ is odd, we can find a generic solution to 
$c_1\left(V_{A,a}\otimes \cn_0\right)=0$ 
by taking $a=1$ and 
\eqn\gensol{
c_1(\cn_0)=\alpha - {n+1\over 2}L.}
Since our aim is to describe $E_6\rightarrow E_7$ transition, we 
start with $n=3,\ a=1$, therefore the heterotic data is specified by a 
rank 3 bundle $U_3=V_3\otimes\pi^*\co_B(\alpha - 2L)$. 
After transition we have a rank two reflexive sheaf $V_2$ related to 
$V_3$ by the exact sequence \exseqS.\ Then it is harder to find generic 
solutions to $c_1(V_2\otimes \cn_0)=0$. 
In fact, it is clear that such a bundle exists if and only if 
\eqn\restriction{
L\equiv 0 \qquad \hbox{mod}\ 2.}
This condition is rather restrictive, but it is satisfied if the 
base of the elliptic fibration is for example $B=\IF_0$ or $B=\IF_2$.
If \restriction\ is not satisfied, we will have to understand more
complicated transitions which change $a$ as well as $n$. This 
is an interesting problem, but we will not address it here.

Let us assume for now that $L=2L^\prime$, \ie\  the anticanonical 
line bundle $\cl$ has a square root $\cl^\prime$.
Then the heterotic data after the transition will be specified 
by the reflexive sheaf 
$U_2=V_2\otimes \co_B\left(\alpha - 3L^\prime\right)$
The exact sequence \exseqS\ implies 
\eqn\exseqL{
0\ra {\cal J}_{p_{01}+\ldots +p_{0k}}\otimes 
\cl\ra U_3\ra U_2\otimes {\cl^\prime}^{-1}
\ra 0.}

To conclude, it might be helpful to express the bundle data in the 
language of \FMWb.\ Using the relations, \rel\ a straightforward 
computation shows that the solution \gensol\ is equivalent to the 
choice 
\eqn\choice{
\lambda = {1\over 2}}
in the notation of \FMWb. The number $\lambda$ determines the 
third Chern class of the heterotic bundle $U_n$. More precisely, 
for a smooth generic bundle we have 
\nref\curio{G. Curio, ``Chiral
matter and transitions in heterotic string
models,'' Phys. Lett. {\bf B435} (1998) 39, hep-th/9803224.}%
\curio\
\eqn\thirdA{
c_3(U_n)=2\lambda \alpha(\alpha-nL).}
This formula is not valid for the rank two reflexive sheaf $U_2$ 
constructed above. As detailed in appendix A, the third Chern class 
of $U_2$ is 
\eqn\thirdB{
c_3(U_2)=2(1+\lambda) \alpha(\alpha-2L).}
Note that $\alpha(\alpha-2L)$ is precisely the number of points on $Z$ 
where $U_2$ is not locally free, \ie\ the number of blowups required 
in the previous construction. This shift in the third Chern class 
is related to the presence of a nonperturbative sector after the 
transition. According to the evidence known so far 
\dgm,\ the reflexive sheaf $U_2$ defines a nonsingular $(0,2)$ CFT. 
However, the complete set of data specifying the heterotic 
vacuum also includes a nonperturbative sector consisting 
of massless excitations localized along certain three dimensional 
space-time defects. These defects are associated to the ideal sheaves 
${\cal J}_{p_{0i}}$ and contribute $-2\alpha(\alpha-2L)$ to the third Chern
class. This is quite similar to the small instanton transition in six 
dimensions, where the fivebranes contribute to the second Chern class
of the gauge bundle. The main difference is that in the present case 
there is no Coulomb branch; the defects cannot leave the Horava--Witten 
boundary.  

\bigskip

\noindent
{\bf Acknowledgments}

We would like to thank Paul Aspinwall, Victor Batyrev, 
Jacques Distler, Antonella Grassi, Mark Gross, 
Ken Intriligator, Andreas Karch, and Eric Sharpe for useful discussions 
and correspondence. DED would like to thank the Mathematical Institute, 
Oxford University and the Center for Geometry and Theoretical Physics, Duke 
University, for hospitality during the completion of this work. GR wishes
to thank the Mathematical Institute, Oxford University for hospitality during
the early stages of this work. The work of DED was supported in part by DOE
grant DE-FG02-90ER40542. The work of BF was supported in part by NSF grant
PHY-9511632 and by a John A.\ Wheeler fellowship. 
The work of DRM was supported in part by the Institute
for Advanced Study, and by NSF grants DMS-9401447
and DMS-0074072.  The work of GR was supported in part by NSF grant
PHY-0070928 and by the Helen and Martin Chooljian Foundation.

\appendix{A}{Chern Classes}

Here we compute the Chern character of the K theoretic direct images 
$f_{!}p_{!}\vp_3$, $f_{!}p_{!}\vpp_3$ using the 
Grothendieck--Riemann--Roch (GRR) theorem. Recall that we have a basic 
diagram 
$$\xymatrix{
 Z_1 \ar[r]^{g} \ar[d]_{p} & \wz \ar[d]^{q}\\
 Z_2 \ar[r]^{f}            & Z. \\
}$$
In order to apply the GRR theorem, we need to know the Chern 
classes of the manifolds $\wz,Z_1,Z_2$ in terms of those of $Z$ 
and the classes of the exceptional divisors. The general formula 
\nref\chern{I.R. Porteous, ``Blowing up Chern classes,'' Proc. Camb. 
Phil. Soc. {\bf 56} (1960) 118.}%
\chern\ is as follows. 

Let $X$ be a projective 
variety and $Y\subset X$ be a smooth subvariety with normal bundle 
$\cal E$. Let $f:X^\prime\ra X$ be the blowup of $X$ along $Y$ and let 
$j:D\hookrightarrow  X^\prime$ be the exceptional divisor. 
Let $g:D=\IP{\cal E}\ra Y$ denote the standard projection.
We define a class 
$v\in H^2(D)$ by 
\eqn\blowclass{
v=j^*j_*({\bf 1}_D).}
Then we have the formula 
\eqn\chernA{
f^*\ch(X)-\ch(X^\prime)=j_*\left[\left(g^*\ch{\cal E}-e^v\right)
\left({1-e^{-v}\over v}\right)\right].}

Now let us summarize the notations and conventions for the above
diagram. 
\halign{\qquad   $#$ \qquad & # \hfill \cr
q:\wz\ra Z & The blowup of $Z$ along a
smooth elliptic fiber of $Z$.\cr
D &  The exceptional divisor of $q:\wz\ra Z$.\cr
E &  The proper transform of the elliptic fiber;
 $D=E\times \IP^1$, $N_{D/\wz}=\co_{D}(-E)$.\cr
g:Z_1\ra \wz & The blowup of $\wz$ along the exceptional 
fiber $\IP^1\times p_{0}$,\cr
  &  where $p_{0}$ is the origin of $E$.\cr
D_{1} & The  exceptional divisor of $g$; 
$D_{1}\simeq \IF_1$. \cr
l & The  proper transform of $\IP^1\times p_{0}$;
$l$ is the negative section of $D_{1}$.\cr
f & The fiber class of $D_{1}$.\cr 
h=l+f &  The positive section of $D_1=\IF_1$;
         $N_{D_{1}/Z_1}=\co_{D_{1}}(-h)$.\cr
D^\prime & The proper transform of $D$; 
$g^*D=D^\prime + D_1$.\cr
E^\prime & The proper transform of $E$, \ie\ 
$g^*E= E^\prime +f$;
$N_{D^\prime/Z_1}=\co_{D^\prime}(-E^\prime-l)$.\cr
p:Z_1\ra Z_2 & Contraction along the $\IP^1$ fibers of 
$D^\prime$. \cr
f:Z_2\ra Z & The blowup of $Z$ at the point 
$p_0=E\cap \sigma$. \cr 
D_2=\IP^2 & The exceptional divisor of $f:Z_2\ra Z$; 
$D_2=p_*D_1$.
\cr}

Using \chernA,\ we compute 
\eqn\toddA{\eqalign{
\td(\wz)&=1-{D\over 2}+\td_2(Z)\cr
\td(Z_2)&=1-D_2+\td_2(Z)-{h\over 3}\cr
\td(Z_1)&=1-\left(D_1+{D^\prime\over 2}\right)+\td_2(Z)-{h-l\over 3}\cr
&=1+\td_1(Z_2)-{D^\prime\over 2}+\td_2(Z_2)+{l\over 3}\cr}}
Let $x$ be a K theory class in $K(Z_1)$. Then we have 
\eqn\grrA{\eqalign{
&\ch_0(p_{!}x)=\ch_0(x)\cr
&\ch_1(p_{!}x)=p_*\ch_1(x)\cr
&\ch_2(p_{!}x)=p_*\left[\ch_2(x)-{D^\prime\over 2}\ch_1(x)\right]\cr
&\ch_3(p_{!}x)=p_*\left[\ch_3(x)-{D^\prime\over 2}\ch_2(x)-{l\over 6}
\ch_1(x)\right]\cr}}
Now let $y$ be a K theory class in $K(Z_2)$. Then 
\eqn\grrB{\eqalign{
&\ch_0(f_{!}y)=\ch_0(y)\cr
&\ch_1(f_{!}y)=f_*\ch_1(y)\cr
&\ch_2(f_{!}y)=f_*\left[\ch_2(y)-D_2\ch_1(y)\right]\cr
&\ch_3(f_{!}y)=f_*\left[\ch_3(y)-{D_2}\ch_2(y)-{h\over 3}\ch_1(y)
\right]\cr}}
Given these formulae, one can compute the 
topological invariants of the K theory classes $f_{!}p_{!}\vp_2, 
f_{!}p_{!}\vp_3, f_{!}p_{!}\vpp_3$. 
The starting point is the spectral cover construction on $\wz$. 
In general, this produces a smooth rank $n$ bundle with Chern character 
\eqn\chernD{\ch(\wv_n)=e^{-\wal}\left({1-e^{(a+n)L}\over 1-e^L}\right)-
{1-e^{aL}\over 1-e^L}+e^{-\wsig}(1-e^{-\wal}).}
Recall that $\wal,\wsig$ are the proper transforms of $\alpha,\sigma$, 
\eqn\proptrans{
\wal=q^*\alpha-D,\qquad \wsig=q^*\sigma.}

The final result can be conveniently expressed in terms of the 
Chern characters of the generic smooth bundle \chern. More precisely, 
let us define $\ch(n,a)$ by the formula
\eqn\defch{
\ch(n,a)=
e^{-\alpha}{1-e^{(a+n)L}\over 1-e^L}-
{1-e^{aL}\over 1-e^L}+e^{-\sigma}(1-e^{-\alpha}).}
Then we find (for $a=1,k=0$)
\eqn\finresA{\eqalign{
&\ch(f_{!}p_{!}\vp_2)=\ch(2,a)+w_Z\cr
&\ch(f_{!}p_{!}\vp_3)=\ch(3,a)+w_Z\cr
&\ch(f_{!}p_{!}\vpp_3)=\ch(3,a),\cr}}
where $w_Z$ is the fundamental class of $Z$.  
An useful intermediate result is
\eqn\interm{
\ch\left(f_{!}p_{!}j_*\left(\co_{D_1}(-(m+1)f)\right)\right)=
-mw_Z.}

\listrefs
\end